\documentstyle[times,pramana,epsf,floats,graphics]{ias}
\def \ie{{\sl i.e.\/}}
\def \vec#1{{\mathbf #1}}
\def \lqcd{\Lambda_{\scriptscriptstyle QCD}}
\def \alphas{\alpha_{\scriptscriptstyle S}}
\def \bose{n_{\scriptscriptstyle B}}
\def \beq{\begin{equation}}
\def \eeq{\end{equation}}
\def \beqa{\begin{eqnarray}}
\def \eeqa{\end{eqnarray}}

\begin{document}
\mark{{QGP}{S.\ Gupta}}
\author{The Quark Gluon Plasma: lattice computations put to experimental test}
\author{Sourendu Gupta}
\address{Dept.\ of Theoretical Physics, Tata Institute of Fundamental
Research, Homi Bhabha Road, Mumbai 400005, India.}
\pacs{11.15.Ha, 12.38.Gc, 12.38.Mh \hfill TIFR/TH/03-02}
\abstract{I describe how lattice computations are being used to extract
experimentally relevant features of the quark gluon plasma. I deal
specifically with relaxation times, photon emissivity, strangeness yields,
event by event fluctuations of conserved quantities and hydrodynamic
flow. Finally I give evidence that the plasma is liquid-like rather in
some ways.}

\maketitle
\section{Introduction}

\begin{flushright}
{\small\sl
 The quagma engineers? That huge ugly brown thing we saw? That was one
  of them?\hfil\\
  {\rm Gregory Benford\/}, in ``Around the curve of a cosmos''}
\end{flushright}

QCD has been tested at zero temperature by its predictions for ``hard
processes'', \ie, processes in which all relevant scales are much
larger than the intrinsic scale, $\lqcd$. This convenience is due to
asymptotic freedom in QCD; at scales much larger than $\lqcd$ the coupling
$\alphas$ is small. At finite temperature, $T$, the scale relevant to
most thermodynamic variables is of order $T$. Since $T_c/\lqcd=0.5$ for
QCD with two light flavours of quarks \cite{precise}, at experimentally
accessible temperatures $T/T_c\sim$ 1--3, the scales are comparable
to $\lqcd$, $g\equiv\sqrt{4\pi\alphas}={\cal O}(1)$, and one deals
with soft physics \cite{expg}. Perturbation theory may remains a rough
guide to intuition. However, since it is sensitive to the infrared, \ie,
non-perturbative length/mass scales, its domain of applicability really
is $g\ll 1$, \ie, $T\ge10^9T_c$. As a result, lattice gauge theory is
the only theoretical tool of direct relevance to experiments currently
being performed at the Relativistic Heavy-Ion collider (RHIC) at the
Brookhaven Lab.

Until recently, the agreement of the energy density at freeze-out in
relativistic heavy-ion collisions with that predicted at $T_c$ by lattice
computations, and the connection between Debye screening and $J/\psi$
suppression, have been the main points of contact between fundamental
QCD computations and experiments. In this talk I will concentrate on
other comparisons, all potentially precise confrontations of lattice QCD
predictions against experiments.  Many of these have emerged in the last
few years and are therefore less well-known. Specifically, I will deal
with predictions of strangeness yields, event to event fluctuations of
conserved quantities, extraction of the speed of sound from the centrality
dependence of elliptic flow and the first estimates of relaxation times
and photon/dilepton production rates. A secondary motive for this talk
is to identify the ways in which thermal perturbation theory may guide
our thinking even in the domain where it is not expected to work.

For $T\ll T_c$ strongly interacting matter is in the confined
phase. Chiral symmetry is spontaneously broken, with pions emerging as
pseudo-Goldstone boson. Since the Dirac operator for quarks has nearly
vanishing eigenvalues, accurate lattice computations are hard. In this
range of temperatures it may be much easier to use effective theories
such as chiral perturbation theory to extract quantities of interest
to experiments. Interesting predictions exist for a lukewarm pion gas
\cite{son} and for the phases of cold and dense QCD \cite{dense}. At
this time it seems that the role of lattice computations is to validate
and determine some of the crucial inputs into such models. A discussion
of this lies outside the scope of this talk.

QCD matter undergoes a phase transition, or at least a rapid cross over at
$T=T_c$. This was the region on which the earliest lattice computations
concentrated--- successfully extracting $T_c$ with high precision,
and estimating the order of the phase transition \cite{oldlat}. The
universality class of the phase transition in the chiral limit still
remains to be reliably extracted--- the main problem here is that
extracting physics at small quark masses requires very large lattices,
thus pushing up the time required to perform accurate numerical lattice
computations. This region of temperature remains of great interest,
since the transition from quarks to hadrons stamps the physics of this
region onto many observables studied at RHIC. Since highly accurate
lattice computations for this region are still underway, this talk will
touch only briefly on this.

Most of the material in this talk is of relevance to the physics of
the temperature range $1.5\le T/T_c\le3$, where $g={\cal O}(1)$, and
the perturbative and non-perturbative scales cannot be separated. As
a result, perturbation theory cannot be numerically accurate and
lattice computations are essential to extract the physics of the plasma.
This talk is divided into four main sections. We begin by an examination
of the quasiparticle modes of the plasma, which allows us to test
perturbative expansions in a theoretically clean setting. The next
two sections concentrate on two thermodynamic quantities of direct
relevance to experiments--- the equation of state and quark number
susceptibilities. The following section is devoted to off-equilibrium
phenomena such as relaxation times, electrical conductivity and photon
(and dilepton) production rates in the plasma.

\section{Perturbation theory: is the QCD plasma a quark gluon plasma?}

\begin{flushright}\narrower\narrower
{\small\sl
 ... quagga, extinct African wild ass like the zebra. quagma, hypothetical
 matter made up of quarks and gluons. quahog, type of edible clam. ...\hfil\\
 {\rm http://phrontistery.50megs.com}, ``List of unusual words''.}
\end{flushright}

\begin{figure}
\begin{center}\scalebox{0.4}{\includegraphics{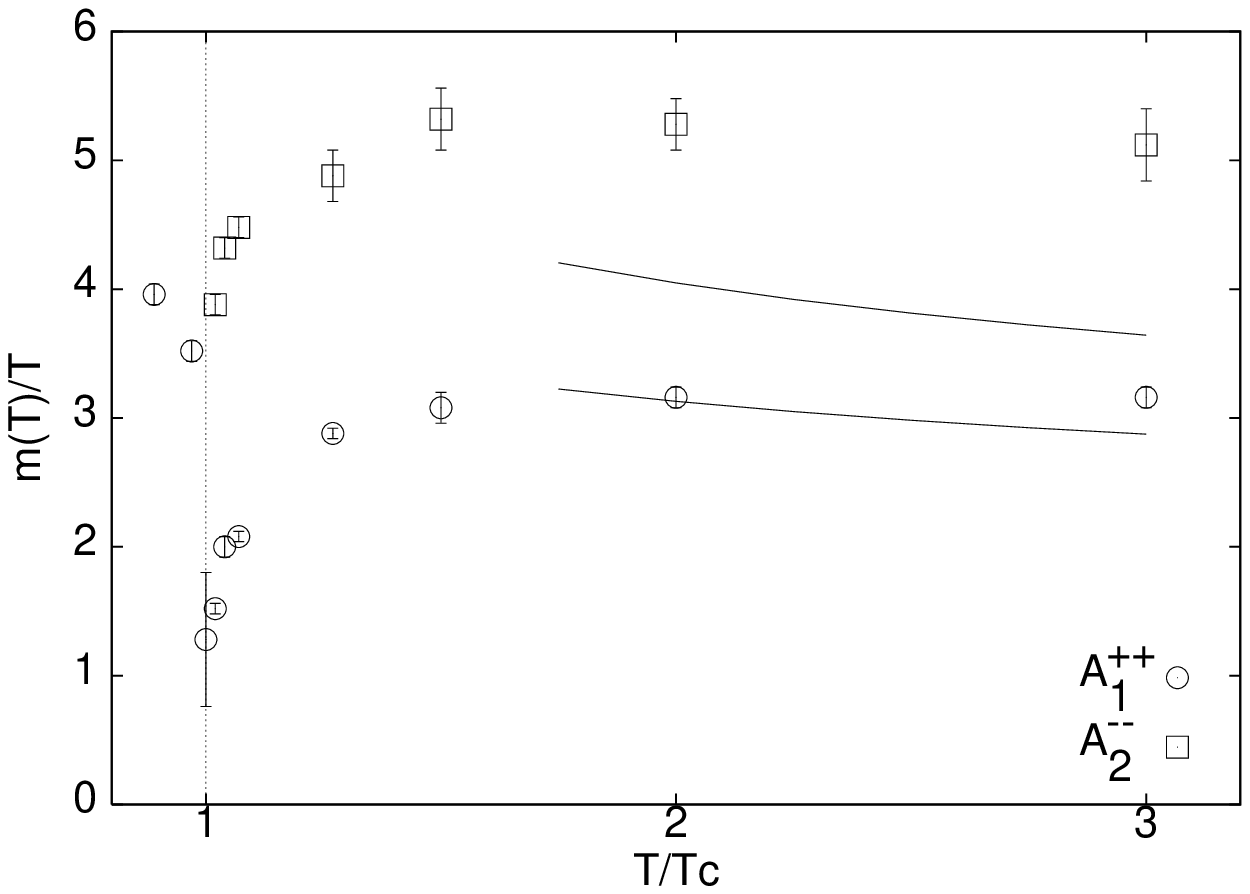}}
              \scalebox{0.4}{\includegraphics{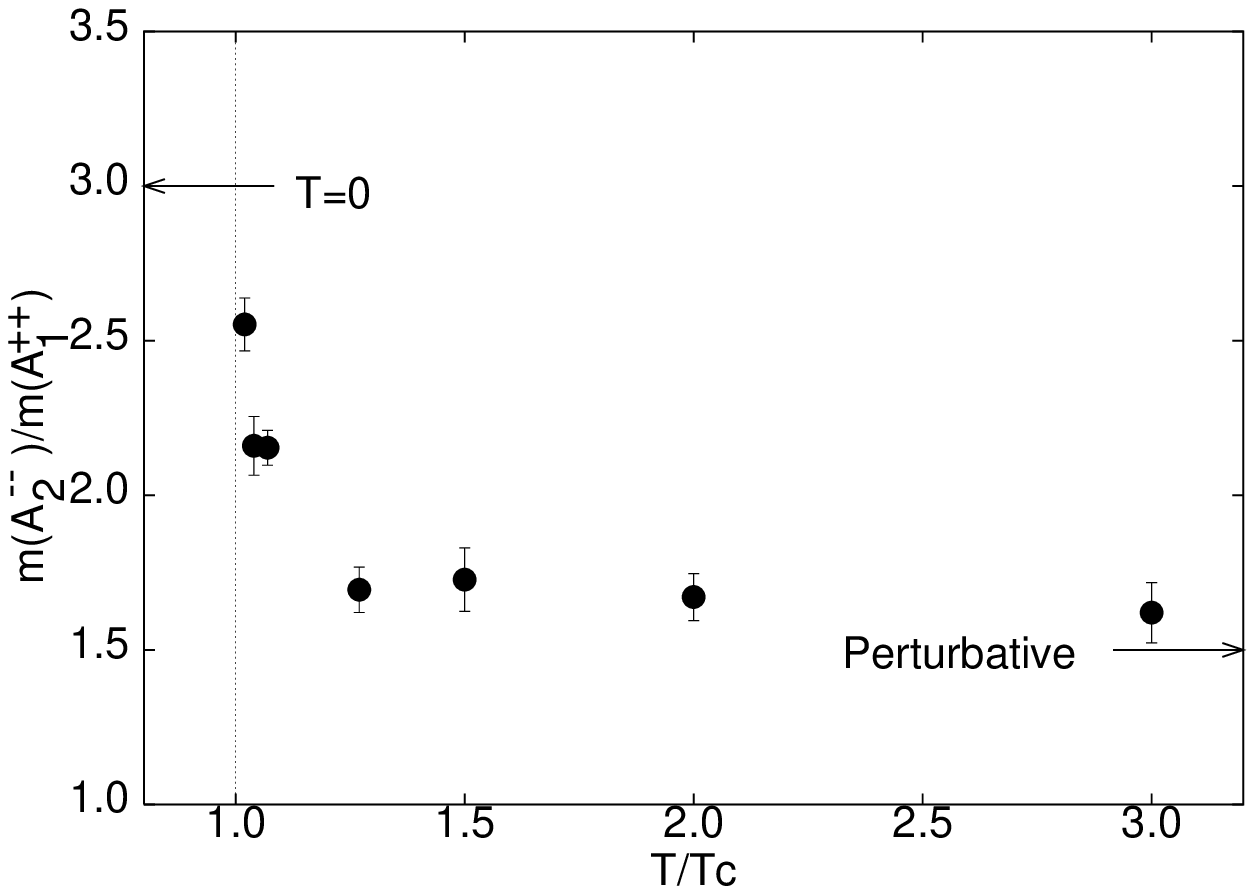}} \end{center}
\caption{The screening masses in quenched QCD from correlations of the two
   operators in eq.\ (\protect\ref{ops}) on the left and their ratio on the
   right. The band is the one-sigma error band on the perturbative prediction
   along with an estimate of the non-perturbative terms from a numerical
   computation in a DR theory \protect\cite{mdeb}.}
\label{fg.saumen}\end{figure}

Perturbation theory is an expansion of the free energy of QCD in
a series in $g$, and is effectively an expansion in terms of gluon
and quark fields. One of the most basic quantities in Euclidean high
temperature perturbation theory is the Debye screening mass. At leading
orders in the perturbative series this has contribution only from the
electric polarisation of the gluon \cite{pols}, however at higher
orders magnetic polarisations also contribute \cite{nadkarni} and,
as a result, the perturbation expansion breaks down at finite order
\cite{arnold}. Perturbative predictions for the Debye screening mass
do not exist close to $T_c$, and lattice studies of Debye screening can
give no meaningful test of perturbation theory \cite{olaf}. A couple of
more limited tests are possible.

The first is to check whether a ``constituent'' gluon picture works
\cite{bernd}. Correlations of the operators
\beq
   A_1^{++} = {\cal R}e\,{\rm Tr}\,L \qquad{\rm and}\qquad 
   A_2^{--} = {\cal I}m\,{\rm Tr}\,L
\label{ops}\eeq
($L$ is the Polyakov line operator, \ie, the flux due to a static quark)
are obtained by two and three electric gluon exchanges to leading order.
If this continues to be true in some sense non-perturbatively, then the
screening masses obtained in these two channels should be in the ratio
3/2. A recent lattice computation (see Figure \ref{fg.saumen}) shows that
this is actually true in the range $1.25\le T/T_c\le3$ \cite{saumen}.
However, detailed studies of other screening masses on the lattice
show that no ``constituent'' picture can be built up in the sector of
magnetic gluons \cite{saumenold}. In fact, magnetic Wilson loops have
been shown to confine \cite{bali}. This is consistent with a picture of
an effective theory for finite temperature QCD in which electric gluons
and magnetic glueballs are the degrees of freedom \cite{gpy}. A detailed
model consistent with the lattice data is under investigation \cite{rob}.

The second is to test a systematic reduction of the theory which goes
by the name of dimensional reduction (DR) \cite{dr}. This attempts to
integrate out the high frequency ($\omega\ge2\pi T$) components of the
theory and produce a long distance effective theory. The couplings in
this effective theory are computed at the scale $2\pi T$ and hence
perturbation theory should be fine as long as $\alphas$ is small
enough. However, the effective theory is fairly complicated (probably
confining) \cite{kajantie} and its long distance properties have to be
extracted by a lattice computation. For quenched QCD, the spectrum of
screening masses obtained from DR \cite{owe} agrees with that from the
full theory for $T\ge2T_c$ \cite{talk00}. One such test is shown in
Figure \ref{fg.saumen}.

For physics in thermal equilibrium, it seems fruitful to think of the
quenched QCD plasma above $1.25T_c$ as containing electric gluons. The
magnetic sector seems confined, thus solving the infrared (Linde)
problems of hot perturbative QCD through the non-perturbative mechanism
of generating ``thermal glueballs'' \cite{magscr}. Closer to $T_c$ there
is not even any evidence for electric gluons. QCD with dynamical quarks
may have a quantitative description in terms of gluons only for $T>6T_c$
\cite{obstr,precise}.

\section{Flow and the equation of state}

\begin{flushright}\narrower\narrower
{\small\sl
 When quagma is allowed to cool and expand its binding superforce decomposes into
 four sub-forces. To my surprise, I understood some of this.\\
  {\rm Stephen Baxter}, in ``On the Orion Line''}
\end{flushright}

A clear signal of collective effects in the final state of a relativistic
heavy-ion collision would be hydrodynamical flow. If flow can be
unambiguously identified in experiments, then the equation of state (EOS)
of QCD matter becomes accessible to measurement, since it is an input to
the hydrodynamical equations. The EOS, \ie, the temperature dependence
of pressure ($P$) , energy ($E$) and entropy ($S$) densities, have been
extracted on the lattice in quenched QCD \cite{eosnf0} as well as in QCD
with two \cite{eosnf2} or four \cite{eosnf4} flavours of dynamical quarks.
It is a remarkable lacuna that these EOS has not yet been put through
the machinery of hydrodynamical codes to confront experiments \cite{jane}.

\begin{figure}
\begin{center}\scalebox{0.3}{\includegraphics{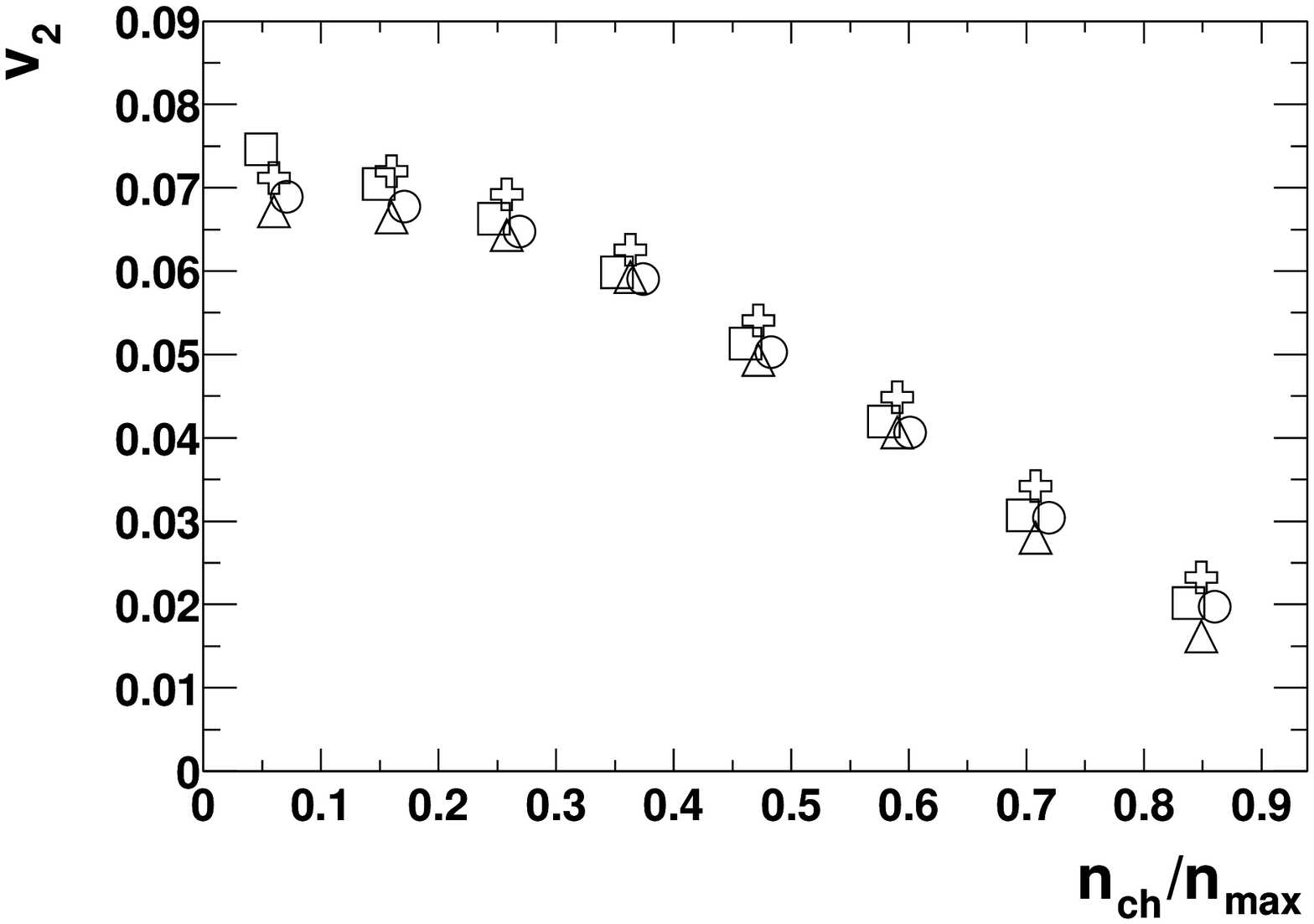}}
              \scalebox{0.43}{\includegraphics{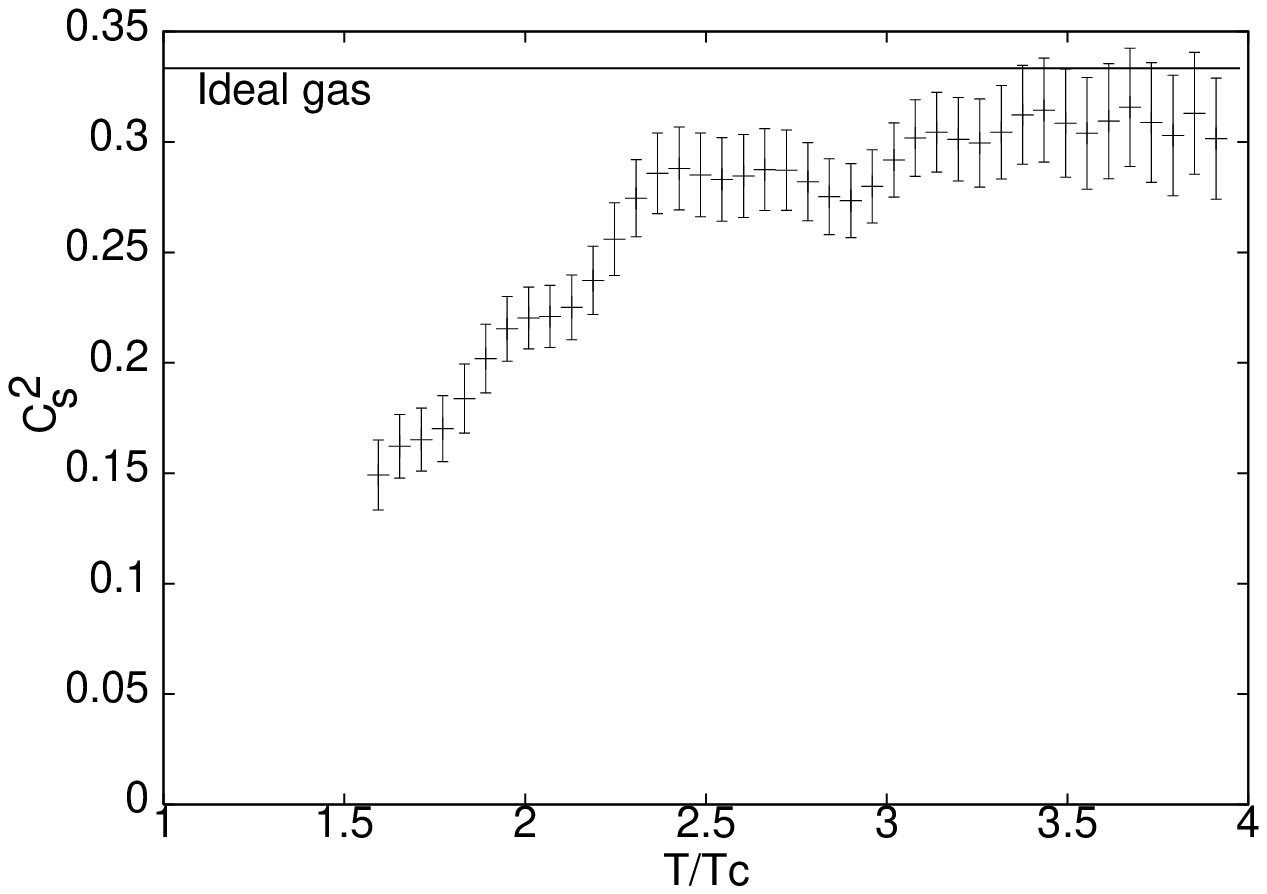}} \end{center}
\caption{On the left are STAR collaboration results on $v_2$ as a function
   of centrality \protect\cite{elliptic}, from the correlation between particle
   pairs consisting of randomly chosen particles (circles), particles with
   opposite sign of charge (crosses), particles with the same sign of
   charge (triangles) and particles with opposite sign of pseudorapidity
   (squares). On the right is a lattice estimate of the speed of sound in
   quenched QCD.}
\label{fg.v2sound}\end{figure}

$P$, $S$ and $E$ deviate from the Stefan-Boltzmann limit strongly near
$T_c$ and by about 20\% even at the highest temperatures at which lattice
computations exist (about $4T_c$). This seems to have no explanation
within perturbation theory, since the perturbative series for $P$
fluctuates wildly as more terms are added; a Borel \cite{borel} or Pad\'e
\cite{pade} summation of the series does not help. Screened perturbation
theory \cite{scrpt} applied to the hard thermal loop resummation does
not produce agreement with the lattice results \cite{eospert}. On the
other hand, there have been reasonably successful attempts to fit the
pressure by a gas of quasiparticles whose masses are the fit parameters
\cite{quasi}. A partially self-consistent resummation also gives good
agreement with the lattice data \cite{eosbir}. More recently the pressure
has been obtained in the DR theory \cite{eosdr}.

Signatures of hydrodynamic flow have been sought in particle spectra and
in HBT radii in the past. At present one of the most promising signals
is elliptic flow \cite{flow}. If hydrodynamics can be trusted, then,
in off-center collisions of two nuclei, the spatial anisotropy leads
to pressure gradients. These drive momentum anisotropies, whose second
Fourier coefficient, $v_2$, is called elliptic flow \cite{v2}. This
has been observed in experiments over a wide range of collider energies
\cite{elliptic}.

At RHIC energies, the variation of $v_2$ with the impact parameter $b$
(which determines the charged multiplicity $n_{ch}$) is claimed to have
a good explanation in terms of hydrodynamic flow \cite{v2hydro}. So
does the variation of $v_2$ with the transverse momenta, $p_t$, of the
particles used to measure it \cite{altv2}. If the initial temperature is
determined independently, then the slope of $v_2$ against $b$ depends
on the speed of sound, $c_s$, since the pressure drives the evolution
of $v_2$. In principle, then, $c_s$ can be measured directly from RHIC
experiments and compared to predictions from the lattice.

Lattice predictions for $c_s$ can be obtained as a byproduct of the
extraction of the EOS. In Figure \ref{fg.v2sound} we show our extraction
of $c_s$ from the data in \cite{eosnf0}. This computation is preliminary
(a more detailed computation is underway), and the main uncertainty is
connected with the fact that the lattice data used have finite lattice
spacing artifacts which need to be compensated for. However, a dip
in $c_s$ near $T_c$ has been seen with two-flavour dynamical quarks
\cite{milc}, and argued to follow from thermodynamic considerations
\cite{thermo}. The most interesting observation is that at the highest
temperatures $c_s$ is close to its ideal gas value, although both $P$
and $E$ are far from ideal. This has also been seen with two flavours
of dynamical quarks \cite{milc}.

\section{Fluctuations, strangeness yields, and quark number susceptibilities}

\begin{flushright}\narrower\narrower
{\small\sl
 We were inducted here by some curious property of the quagma, so I suppose.\\
  {\rm Gregory Benford}, in ``Around the curve of a cosmos''}
\end{flushright}

\begin{figure}
\begin{center}\scalebox{0.7}{\includegraphics{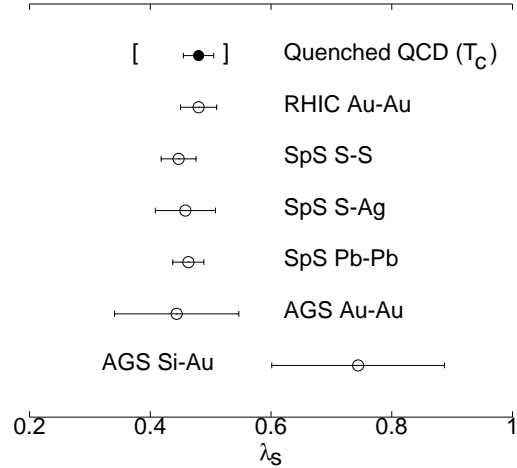}}\end{center}
\caption{The Wroblewski parameter extracted from data \protect\cite{cleymans}
   compared to a quenched lattice computation \protect\cite{ours}. The bars
   are statistical errors. The brackets denote the envelope of uncertainties
   due to extrapolations from $T\ge1.5T_c$ down to $T_c$.}
\label{fg.wrob}\end{figure}

Event by event fluctuations in conserved quantities such as the charge
or baryon number \cite{fluct} are proportional to quark number
susceptibilities
\beq
   \chi_{fg} =\left.-\frac TV
      \frac{\partial^2\log Z}{\partial\mu_f\partial\mu_g}\right|_{\mu_f=\mu_g=0},
\label{qns}\eeq
where $Z$ is the partition function of QCD and $\mu_f$ is the chemical
potential for flavour $f$ \cite{gott}. Further details, including
those of the evaluation of these susceptibilities on the lattice can
be found in several recent reviews \cite{qns}. It is interesting to
note that recent lattice computations \cite{qnslat} for the diagonal
susceptibilities ($\chi_{ff}$) can be reproduced in a skeleton graph
resummation \cite{qnsbir}, dimensional reduction \cite{qnsdr} and also in
a quasiparticle picture \cite{qnsqp}.  The off-diagonal susceptibilities
are found to be zero in lattice computations; there seems to be no
explanation for this in models.

Measured fluctuations \cite{fluce} are thought to be proportional to the
ratio $\chi/S$.  Lattice computations for these are under good control for
$T>T_c$, but the region $T<T_c$ requires more work. Present day lattice
data \cite{qnslat} indicate a hierarchy of fluctuations for baryon number
($\chi_B$), electric charge ($\chi_Q$) and strangeness ($\chi_s$)---
\beqa
\nonumber
  \chi_B<\chi_Q<\chi_s\qquad && \qquad(T>T_c), \\
  \chi_B>\chi_Q>\chi_s\qquad && \qquad(T<T_c).
\label{hier}\eeqa
The inversion of the hierarchy as one crosses $T_c$ may be a possible
experimental signal of the phase transition.

One of the most interesting pieces of information that the lattice can
supply is for the strangeness yield, which is measured very accurately in
experiments, and hence has attracted much attention \cite{strange}. This
yield is parametrised as the Wroblewski parameter, $\lambda_s$, which
is the relative number of primary produced strange to light quarks
\cite{wrob,cleymans}.  Clearly, $\lambda_s$ is the ratio of imaginary
parts of the complex susceptibilities in these flavour channels. Under
reasonable (and testable) assumptions \cite{ours}
\beq
   \lambda_s=\frac{2\chi_{ss}}{\chi_{uu}+\chi_{dd}},
\label{wrob}\eeq
thus allowing us to compute this quantity on the lattice. Results obtained
in quenched QCD \cite{ours} are exhibited in Figure \ref{fg.wrob}.
We expect this ratio to be fairly insensitive to quenching artifacts. A
computation in dynamical QCD with two flavours at $T_c$ is now underway.

\section{Relaxation times, photon emissivity and the electrical conductivity of a plasma}

\begin{flushright}\narrower\narrower
{\small\sl
 Quagma ... was both the Red Dragon and the Green Dragon. It was light and the
 light was good.\\
 {\rm Jonathan S.\ McDermott} in http://caraig.home.mindspring.com/rant020206.html}
\end{flushright}

We turn next to non-equilibrium phenomena in the QCD plasma.  These are of
very direct relevance to heavy ion experiments, since the matter formed in
the fireball is fully out of equilibrium initially. Of interest are limits
on how fast it equilibrates with respect to the strong interactions, how
fast local thermal fluctuations diffuse away, how quickly a hard probe
(such as a jet) loses energy, whether the system remains forever out of
equilibrium in electroweak interactions, and if so, the rate at which
it radiates leptons and photons. Over the last two years perturbation
theory and lattice computations have reached a stage where we can begin
to constrain the answers seriously.

The most crucial piece of information that is required is of the
equilibration time. Hydrodynamic explanations for particle spectra,
HBT radii and, especially, elliptic flow, all require relatively small
equilibration times in the plasma (0.6--1 fm) \cite{ttime}, implying
that transport related cross sections are huge. Experimental evidence
for jet quenching \cite{jetq}, particularly the damping of away-side
jets \cite{jetaway}, are also indicative of small relaxation times or
rapid energy flows. These time scales, or the corresponding transport
coefficients are intimately related to large angle or multiple small
angle (Landau-Pomeranchuk-Migdal, LPM) scattering and are of the order
of $1/g^4\log(1/g)T$ when $g$ is small enough \cite{kinetic}. The
Kubo formul{\ae} relate these transport coefficients to the zero
energy ($\omega=0$) limits of the imaginary parts of certain retarded
correlators. When these correlators are evaluated in perturbation theory,
the multiparticle states which contribute to it have momenta $(k^i_0,\vec
k^i)$ which sum up to zero ($i$ labels particles). However, when these
intermediate states are massless, each of the $k^i_0\simeq\omega$ can be
zero. Then while integrating over $k^i_0$, the contour is pinched between
these poles. Interactions, specifically the transport cross sections,
throw these poles slightly off-axis, but the pinch still gives a bump in
the imaginary part of the correlators. The effect of such bumps, which
are seen to persist beyond the pertubative regime, is to give rise to
transport coefficients \cite{aarts}.

The simplest of this class of problems deals with electromagnetic
interactions.  The transport coefficient one deals with is the ohmic
conductivity, $\sigma$, \ie, the response of the QCD plasma to an external
static and spatially uniform electric field, $E$. The result of applying
such a field is to set up a current $j=\sigma E$ in the direction of the
field. A Kubo formula relates $\sigma$ to the imaginary part, $\rho$,
of the retarded current-current correlator in equilibrium---
\beq
   \sigma(T) = \frac16\left.\frac{\partial}{\partial\omega}
      \rho_i^i(\omega,{\vec p=\vec0},T)\right|_{\omega=0},
\label{cond}\eeq
where all spatial components $i$ are summed over. There is a finite and
non-vanishing ohmic conductivity as long as $\rho_i^i$ is linear near
zero energy. The photon emissivity is given by
\beq
   \omega\frac{d\Omega}{d^3\vec p} = \frac1{8\pi^3}
      \bose(\omega,T) \rho_\mu^\mu(\omega,\vec p,T),
\label{rate}\eeq
where $\Omega$ is the number of photons produced per unit volume per
unit time. This is equal to the observed photon rate if the reabsorption
rate is very small--- in which case the medium is out of equilibrium
with respect to the EM coupling $\alpha$. In this work we shall take
$\omega={\mathbf p}=0$, and hence obtain the soft photon production
rate. Since $\rho_{00}=0$ for $\vec p=0$, the soft photon rate can be
obtained once $\sigma$ is computed. Extracting $\rho_i^i$
from lattice computations needs the maximum entropy method \cite{mem}
or other Bayesian techniques \cite{sigma}.

\begin{figure}
\begin{center}\scalebox{0.6}{\includegraphics{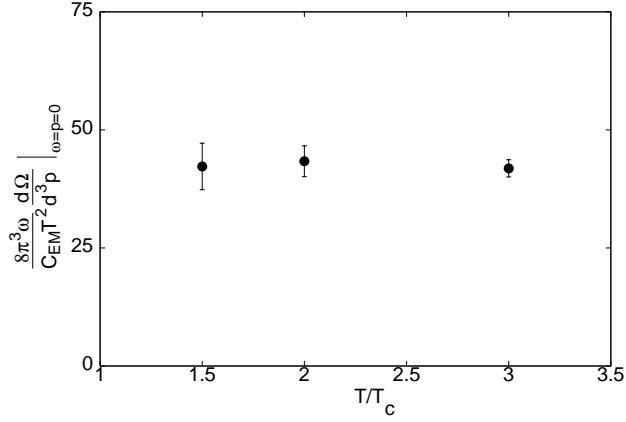}}\end{center}
\caption{The scaled soft photon emissivity obtained from a lattice computation
   \protect\cite{sigma}. The quantity on the abscissa is equal to $6\sigma/T$.}
\label{fg.photon}\end{figure}

The soft photon production rate from the plasma phase of hadronic matter
has long been of importance to searches for the QCD phase transition,
especially due to persistent observations of enhancements in heavy-ion
collisions over proton-proton rates \cite{wa98}. Consequently, there has
been a long history of attempts at perturbative computations of this
rate \cite{history}. The first lattice computation in quenched QCD of
dilepton (off-shell photon) rates \cite{dilepton} showed good agreement
with perturbative results for $\omega>3T$.  Recently the leading order
computation of the photon production rate was completed \cite{amy}.
For the transport coefficient one has $\sigma\propto \alpha T/g^4\log
g^{-1}$, to leading-log accuracy, with a known proportionality constant
\cite{amy2}.  The first computation of $\sigma$ and hence of the soft
photon emissivity from a quenched lattice computation has now been
performed for $1.5\le T/T_c\le3$ \cite{sigma}.  It turns out that
\beq
   \frac\sigma T\approx 7 C_{EM},\quad({\rm for\ }1.5\le T/T_c\le3)\quad
   {\rm where}\quad C_{EM}=4\pi\alpha\sum_f e_f^2,
\label{value}\eeq
and $e_f$ is the charge of a quark of flavour $f$. The corresponding
soft photon emissivity is shown in Figure \ref{fg.photon}. Clearly,
for fireball dimensions less than $1/\sigma=1/7C_{EM}T\approx3$ fm,
the plasma is transparent to photons and this emissivity is also the
detection rate of photons.

The diffusion coefficient of quarks can also be obtained in the same
computation using the Einstein relation $\sigma=4\pi\alpha\sum_f
e_f^2\chi_{ff} D_f$ ---
\beq
   T D_f = \left(\frac{T^2}{\chi_{ff}}\right)\,\left(\frac\sigma{C_{EM}T}\right),
\label{diffu}\eeq
where $\chi_{ff}$ is the quark number susceptibility defined in eq.\
(\ref{qns}) \cite{amy2}.  A characteristic relaxation time, $\tau_R$,
is the time for quarks for diffuse a distance equal to the screening
length $1/T$. Then, we have
\beq
   \tau_R \approx \frac1{DT^2}\approx\frac1{7T}.
\label{relax}\eeq
For $1.5\le T/T_c\le3$ this is much smaller than a fermi. However,
the relaxation time for charge carries an extra power of $\alpha$ in
the denominator and hence is two orders of magnitude larger.  This is
the reason why charge fluctuations may be detectable.

The relaxation time required in jet quenching has to do with the
gluon-dominated transport coefficient $\hat q$, which measures momentum
transport transverse to the external force \cite{transq}.  This transport
coefficient remains to be measured on the lattice, but there is no
reason to suspect that it leads to a significantly longer relaxation
time. A complete theory of equilibration does not exist at this time
\cite{thermal}, but given such small relaxation times near equilibrium,
it does not seem implausible that equilibration times are also small.

On purely phenomenological grounds it is clear that extremely fast
thermalization and jet quenching is not compatible with a fireball that
is very transparent to photons. The ratio of the relevant scales is just
$C_{EM}\approx1/20$. If the former scale is about 0.1--0.15 fm, then the
latter scale must be in the range 2--3 fm. Thus, the fireball produced
at RHIC is marginally transparent to soft photons, whereas the larger
expected size of a fireball at LHC would only allow photon detectors to
look 2--3 fm inside the surface of the fireball.

\section{(not the) Conclusion}

\begin{flushright}\narrower\narrower
{\small\sl
   ... the Earth is a type-13 civilization. Type 13 civilizations destroy
   themselves by turning their planet into degenerate matter looking for
   the Higgs boson.\\ {\rm murphy@panix.com} posted on Slashdot}
\end{flushright}

Let me introduce a dimensionless parameter which classifies several aspects
of the physics that I have been talking about--- the liquidity, defined by
\beq
   \ell = \tau S^{1/3} \approx \tau E^{1/4},
\label{liquid}\eeq
where $\tau$ is the transport mean free time. The non-relativistic
analogue of $S$ is the number density, so that $\ell$ is the mean free
path in units of the interparticle spacing. For gases we expect this
number to be large.  A liquid would be characterised by values of $\ell$
close to unity.

In the perturbative expansion, when $g\ll1$, we have $S\simeq T^3$,
$\tau\simeq1/Tg^4\log(1/g)$, and hence $\ell\simeq1/g^4\log(1/g)\gg1$.
As a result, perturbation theory describes only the dilute, gaseous,
phase of the QCD plasma. In experiments one finds $E\simeq 1$
Gev/fm${}^3$ and $\tau<1$ fm, giving $\ell<1.5$, and matter that is
definitely liquid.  We shall continue to call this phase a plasma, in
view of the screening phenomena that occur (but remain to be rigorously
demonstrated in experiments). However it is important to remember that
transport coefficients are dominated by interactions, as in liquids,
and not by long mean free paths, as in gases.  The lattice studies now
seem to indicate liquid-like behaviour for $T\le3T_c$, thus bringing us
closer to an interpretation of heavy-ion collisions as quark matter.

The departure of $c_s^2$ from its gas value for $T<2T_c$ and the rapid
fall in $S$, also indicate that the plasma changes character in the
temperature region 2--3$T_c$. However, there is no evidence of a phase
transition between the gaseous and liquid like extremes of the QCD
plasma. This is likely to be the reason that perturbative expansions
around some quasi-particle pictures give a qualitative description of
static quantities such as $S$, $E$ or $\chi$, not far from $T_c$. However,
the experimental numbers indicate that this is unlikely to be the case
for dynamics.

Liquid-like behaviour means that dissipative effects are important to
the fluid dynamics--- in the relation between the HBT, single particle
spectra and elliptic flow. In addition, the supersonic motion of jets
through the liquid should give rise to many interesting colour-MHD effects
apart from jet quenching.  One near-term target for the lattice theory is
to estimate the various transport coefficients and thereby determine the
relative efficiency of various physical mechanisms for entropy production.

It is a pleasure to thank my collaborators, Saumen Datta, Rajiv Gavai,
and Pushan Majumdar, for discussions. I would also like to thank Gert
Aarts, Jean-Paul Blaizot, Francois Gelis, Jean-Yves Ollitrault, Toni
Rebhan and Jose Resco for communications and discussions. Our lattice
computations are largely performed on Compaq Alphas of the Department
of Theoretical Physics, TIFR.


\begin{thebibliography}{99}
\bibitem{expg}
   At $T=0$ the perturbative expansion is in powers of
   $\alphas=g^2/4\pi$. Stronger infrared divergences for $T>0$ causes
   this perturbative expansion to break down. The correct expansion is
   in powers of $g$ and $\log(1/g)$. The sensitivity to non-perturbative
   scales is the Linde problem \cite{linde}.
\bibitem{pols}
   Since the finite temperature theory is defined in a chosen frame, it makes
   sense to talk of electric and magnetic polarisations separately. They can
   also be given covariant meanings.
\bibitem{magscr}
   An old argument for the absence of magnetic screening \cite{dr} is
   worth repeating.  Electric screening is due to thermal fluctuations
   popping out charged pairs from the vacuum. Since there are no light
   magnetic monopoles to pop out, one does not expect magnetic screening.
\end{thebibliography}
\end{document}